\documentclass[conference]{IEEEtran}

\IEEEoverridecommandlockouts
\def\BibTeX{{\rm B\kern-.05em{\sc i\kern-.025em b}\kern-.08em
    T\kern-.1667em\lower.7ex\hbox{E}\kern-.125emX}}
    
\usepackage{graphicx}
\usepackage{cite}
\usepackage{picinpar}
\usepackage{amsmath}
\usepackage{url}
\usepackage{flushend}
\usepackage[latin1]{inputenc}
\usepackage{colortbl}
\usepackage{soul}
\usepackage{multirow}
\usepackage{pifont}
\usepackage{color}
\usepackage{alltt}
\usepackage{enumerate}
\usepackage{pbox}
\usepackage{fancyhdr}
\usepackage{lineno} 
\usepackage{bm}

\usepackage{amssymb}
\usepackage{stfloats}
\usepackage{algorithmic}
\usepackage[ruled,vlined]{algorithm2e}

\begin{document}

\title{Exploring Blockchain for The Coordination of Distributed Energy Resources}

\author{\IEEEauthorblockN{Qing~Yang}
\IEEEauthorblockA{\textit{College of Electronics and Information Engineering} \\
\textit{and Blockchain Technology Research Center} \\
\textit{Shenzhen University}\\
Shenzhen, China \\
yang.qing@szu.edu.cn}
\and
\IEEEauthorblockN{Hao~Wang*}
\IEEEauthorblockA{\textit{Department of Data Science and Artificial Intelligence} \\
\textit{Faculty of Information Technology} \\
\textit{Monash University}\\
Melbourne, VIC 3800, Australia \\
hao.wang2@monash.edu}
\thanks{*Corresponding author: Hao Wang.}
}
\maketitle

\begin{abstract}
The fast growth of distributed energy resources (DERs), such as distributed renewables (e.g., rooftop PV panels), energy storage systems, electric vehicles, and controllable appliances, drives the power system toward a decentralized system with bidirectional power flow. The coordination of DERs through an aggregator, such as a utility, system operator, or a third-party coordinator, emerges as a promising paradigm. However, it is not well understood how to enable trust between the aggregator and DERs to integrate DERs efficiently. In this paper, we develop a trustable and distributed coordination system for DERs using blockchain technology. We model various DERs and formulate a cost minimization problem for DERs to optimize their energy trading, scheduling, and demand response. We use the alternating direction method of multipliers (ADMM) to solve the problem in a distributed fashion. To implement the distributed algorithm in a trustable way, we design a smart contract to update multipliers and communicate with DERs in a blockchain network. We validate our design by experiments using real-world data, and the simulation results demonstrate the effectiveness of our algorithm. 
\end{abstract}


\section{Introduction}
Various factors, such as concerns about climate change, have driven the fast growth of distributed energy resources (DERs) in the power system. These DERs often include distributed renewables (e.g., rooftop PV panels), energy storage systems, electric vehicles, and controllable appliances, such as heating, ventilation, and air conditioning (HVAC), water heaters, and washers. Due to the stochastic and non-dispatchable nature of DERs, it is challenging to manage DERs by the centralized approach used in the operation of today's power system. A paradigm shift toward a decentralized power system with bidirectional power flow is gaining attention to enable the integration of DERs. The widely accepted roadmap is integrating hundreds of thousands of DERs through an aggregator, such as a utility, system operator, or a third-party coordinator. However, it is not well understood how to build trust between DERs and the aggregator to coordinate DERs. This paper aims to explore blockchain, known as a decentralized ledger, to facilitate a trustable system for the decentralized coordination of DERs.

\subsection{Related works}
A large body of literature studied how to coordinate DERs for various applications. 
For example, the aggregation of controllable loads, distributed generators, and energy storage was studied in \cite{kasaei2017optimal} for mitigating the impact of the variable renewable generations. The integration of distributed renewables as a virtual power plant was studied in \cite{naval2020virtual}.
Other research considered coordinating DERs to provide demand response (DR) in addition to feeding electricity back to the grid. Customer-owned DERs, such as HVAC, water heaters, and cloth washers, can be switched on or off to absorb or shed power. For example, using electric space heating loads in a building to defer power consumption was studied in \cite{thavlov2014utilization}. How to coordinate flexible loads to build demand response capacity was illustrated in \cite{royapoor2020building}.
Recent studies also explored other transactive energy applications for DERs, such as ancillary services \cite{mashhour2010bidding,jioptimal,borne2016provision} and peer-to-peer energy trading \cite{alam2019peer}. 
However, the above studies adopted centralized solution methods, which causes privacy concerns and operational problems, as DERs are often owned by independent prosumers.

Great efforts have also been made to develop distributed solutions to the coordination of DERs. For example, Wang and Huang in \cite{wang2016incentivizing} designed an incentive mechanism and a distributed algorithm for energy trading among interconnected microgrids. Yang and Wang in \cite{yang2020cooperative} developed a distributed transactive energy management scheme for smart homes to energy trading and manage their HVAC systems. You et al. in \cite{you2017scheduling} developed distributed solutions to the scheduling problem of battery swapping for electric vehicles. A fully distributed algorithm was proposed in \cite{chen2018fully} for a virtual power plant considering the transmission limits and constraints of DERs. Most of the developed distributed algorithms shared similar underlying methodologies, such as primal-dual decomposition and alternating direction method of multipliers (ADMM). However, such algorithms often need to be executed by DERs and a central computing node that updates the dual variables. Such a design has a risk of single-point failure and also requires a verifiable and trustable computing environment. This paper is motivated to address the above problem and facilitate a trustable system for the coordination of DERs.

\subsection{Our Work and Contributions}
Blockchain \cite{bashir2018mastering} is a decentralized ledger that can provide trust and transparency to the transactions. Blockchain can also execute smart contracts to run generic programs to enable various functions. Recently, the energy sector has been keen to explore blockchain technology to enable transactional digital platforms and improve the efficiencies of the system. For example, LO3 Energy \cite{exergy} deployed a blockchain system to allow online payments for energy trading in a microgrid. 

In this work, we take a further step to design a blockchain system for DERs to enable distributed optimization and build a trustable and efficient future energy system. Specially, we formulate an optimization problem for DERs to interact with each other and with the grid by feeding back energy and performing demand response. Given the popularity of ADMM in the distributed energy system, we also decompose our DER optimization problem and solve it using ADMM to coordinate DERs in a distributed manner. We design a trustable DER coordination system based on the blockchain technology by implementing the updating rules for the dual variables in a smart contract on the blockchain system. Our design provides a generic approach to the implementation of distributed optimization algorithms on blockchain and thus enable a trustable and efficient system for the coordination of DERs in the future power grid.

The remainder of this paper is organized as follows. Section~\ref{sec:model} introduces the system model and formulates the DERs cost minimization problem. Section~\ref{sec:algorithm} presents the design of the distributed algorithm and blockchain system for the coordination of DERs. Section~\ref{sec:results} presents the simulation results of DER coordination. Section~\ref{sec:conclusion} concludes our work.

\section{System Model and Formulation}\label{sec:model}

\begin{figure}[!t]
    \centering
    \includegraphics[width=8.5cm]{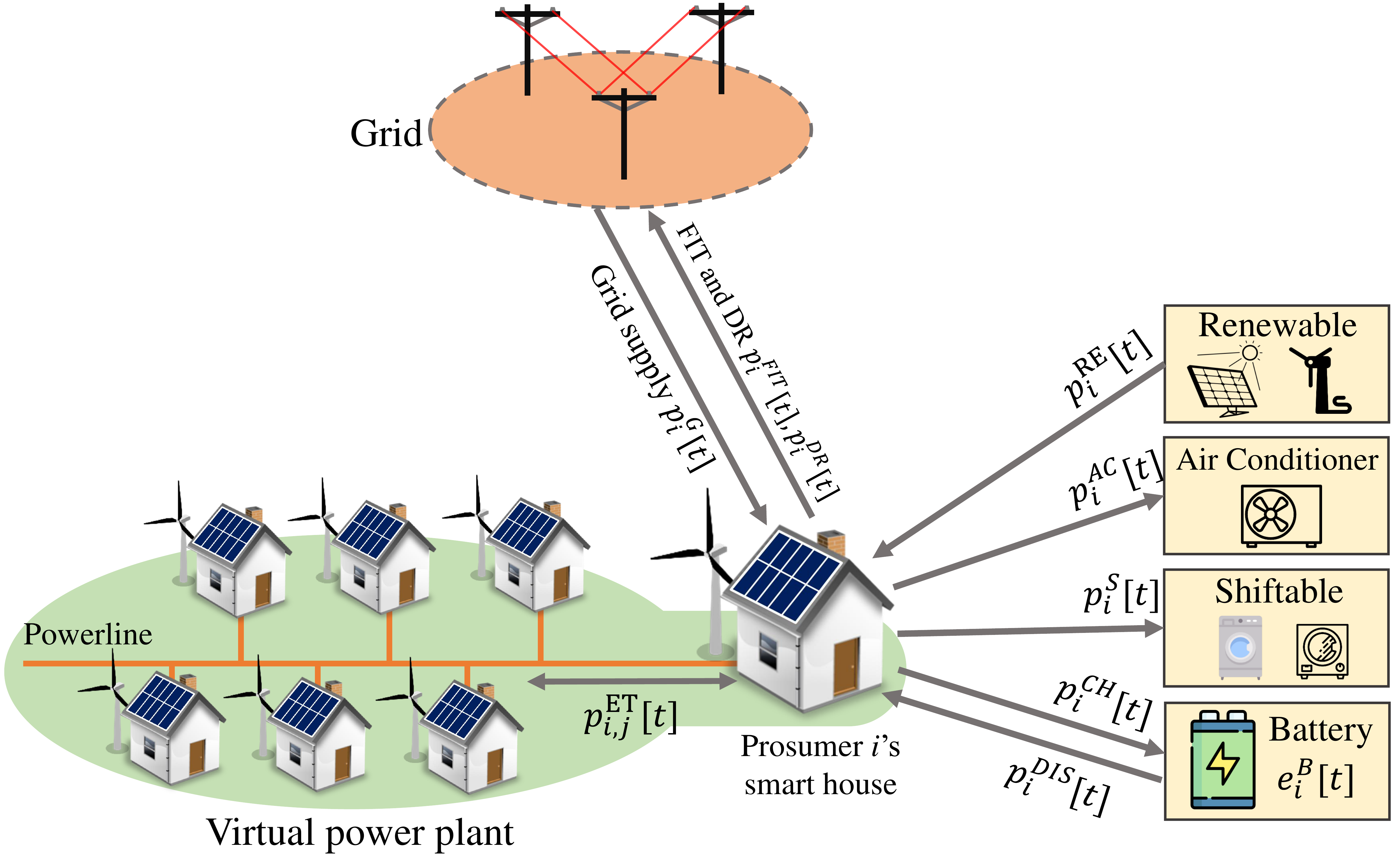}
    \caption{The system model and the operational principle for the coordination of DERs.}
    \label{f:system}
\end{figure}

We consider a group of prosumers denoted as $\mathcal{N} {=} \{1,2,\dots,N \}$, and each prosumer $i {\in} \mathcal{N}$ has a set of DERs, including flexible loads (such as air conditioners and shiftable loads), distributed renewables, and behind-the-meter energy storage, as shown in Fig.~\ref{f:system}. The DERs are coordinated by a blockchain-based coordination system to optimize their energy schedule over a time horizon $\mathcal{T} {=} \{1,2,\dots,T\}$, which is evenly divided into $T$ time slots. For example, in a daily schedule, each day can be divided into $T=24$ time slots.

\subsection{prosumers and Distributed Energy Resources}
Before modeling DERs, we present the prosumers' grid power purchase, as prosumers are interconnected in a distributed network. We denote the power purchase of prosumer $i$ in time slot $t \in \mathcal{T}$ as $p_i^{\text{G}}[t]$, and its grid is written as
    \begin{align}
            \mathcal{C}_i^{\text{G}} = \pi_{\text{E}} \sum\nolimits_{t\in \mathcal{T}} p_i^{\text{G}}[t] + \pi_{\text{D}} \max_{t\in \mathcal{T}} p_i^{\text{G}}[t], \nonumber
    \end{align}
which consists of two charges to the prosumers. The first term $\pi_{\text{E}} \sum\nolimits_{t\in \mathcal{T}} p_i^{\text{G}}[t]$ is called the energy charge, which is based on the accumulative energy consumed. The second term $\pi_{\text{D}} \max_{t\in \mathcal{T}} p_i^{\text{G}}[t]$ is called the demand charge, which is based on the peak demand over the horizon $T$. Such a demand charge drives prosumers to reduce their peak load and thus alleviate the system peak. Given the physical capacity, the electricity purchased from the grid satisfies the following constraints:
    \begin{align}
            0 \leq  p_i^{\text{G}}[t] \leq P_i^{\text{G}}, &~ \forall i \in \mathcal{N}, t \in \mathcal{T}, \label{constraint-load1}
    \end{align}
in which $p_i^{\text{G}}[t]$ is non-negative and bounded by the capacity of the grid power supply $P_i^{\text{G}}$.

In the following, we model DERs, including flexible loads, renewables, and energy storage.

\subsubsection{Flexible Loads}
We take the HVAC as a typical example of an adjustable appliance. 
The indoor temperature of prosumer $i$ in time slot $t$ is denoted as $\tau^{\text{In}}_i[t]$, and $\tau^{\text{Ref}}_i$ is the preferred indoor temperature. The indoor temperature evolves with the HVAC load $p^\text{AC}_i[t]$ and the outdoor temperature $\tau^{\text{Out}}[t]$. Based on the model of HVAC \cite{cui2019,lu2012evaluation}, the dynamics of the indoor temperature is
    \begin{align}
            \tau^{\text{In}}_i[t] = \tau^{\text{Out}}[t] - \left( \tau^{\text{Out}}[t] - \tau^{\text{In}}_i[t-1] \right) \text{e}^{1/RC} \nonumber\\
            + \gamma_i p^\text{AC}_i[t{-}1], \forall i \in \mathcal{N}, \forall t \in \mathcal{T}, \label{constraint-load2}
    \end{align}
in which the coefficients $R$ and $C$ are the equivalent heat capacity and thermal resistance of the HVAC, respectively, and $\gamma_i$ indicates the operating mode of heating or cooling.

We model prosumer $i$'s experience of using HVAC as a
discomfort cost, which measures the deviation of the indoor temperature $\tau^{\text{In}}_i[t]$ to the preferred temperature $\tau^{\text{Ref}}_i$. Specifically, the discomfort cost is a quadratic function shown as
    \begin{equation}
            \mathcal{C}_{i}^{\text{AC}} = \omega_{\text{AC}} \sum_{t \in \mathcal{T}} \left( \tau^{\text{In}}_i[t] - \tau^{\text{Ref}}_i \right)^{2}, ~ \forall i \in \mathcal{N}, \nonumber
    \end{equation}
in which the coefficient $\omega_{\text{AC}}$ is the prosumer's sensitivity to the discomfort. Also, the indoor temperature should be kept within an accepted range, shown as
\begin{equation}
    \underline{T}^{\text{AC}} \le \tau^{\text{In}}_i[t] \le \overline{T}^{\text{AC}}, \forall i \in \mathcal{N}, \forall t \in \mathcal{T}, \label{constraint-load3}
\end{equation}
where $\underline{T}^{\text{AC}}$ and $\overline{T}^{\text{AC}}$ are the upper-bound and lower-bound for the indoor temperature at all times.

Another type of load called shiftable load $p_i^{\text{S}}[t]$ represents the appliances with their usage shifted over a time window $\mathcal{T}_i^{\text{S}}$. Typical shiftable appliances include washers that can be shifted to any available time slots. But prosumers often have a routine schedule to use these appliances, and we denote the preferred schedule for the shiftable load as $P_i^{\text{S}}[t]$. Prosumers can schedule shiftable loads within $\mathcal{T}_i^{\text{S}}$ as long as the task is completed. Therefore, we have the following constraints for shiftable loads:
    \begin{align}
                \sum_{t} p_i^{\text{S}}[t] = \sum_{t} P_i^{\text{S}}[t], \, t \in \mathcal{T}_i^{\text{S}}, ~i \in \mathcal{N}. \label{constraint-load4}
    \end{align}

Rescheduling shiftable appliances change prosumers' routine behaviors and cause discomfort as well. Similarly, we model the discomfort cost as a quadratic function as 
    \begin{align}
            \mathcal{C}_i^{\text{S}} = \omega_{\text{S}} \sum_{t \in \mathcal{T}_i^{\text{S}}} \left( p_i^{\text{S}}[t] - P_i^{\text{S}}[t] \right)^{2}, \nonumber
    \end{align}
in which the coefficient $\omega_{\text{S}}$ is prosumers' sensitivity to load shifting. For both discomfort costs $\mathcal{C}_{i}^{\text{AC}}$ and $\mathcal{C}_i^{\text{S}}$, we use quadratic functions, which are widely used in energy economics literature to capture marginal increase in costs.

\subsubsection{Distributed Renewables}
In addition to purchasing electricity from the grid, prosumers have distributed renewable generations to use. We denote prosumer $i$'s renewable generation in time slot $t$ as $P_i^{\text{RE}}[t]$. Prosumers can choose to use renewable supply to meet their loads and also feed renewable back to the grid. 
The local renewable energy usage $p_i^{\text{RE}}[t]$ of prosumer $i$ should satisfy the following constraints:
    \begin{align}
            0 \leq p_i^{\text{RE}}[t] \leq P_i^{\text{RE}}[t], &~ \forall i \in \mathcal{N}, t \in \mathcal{T}. \label{constraint-load5} 
    \end{align}
Note that we assume that prosumer $i$'s renewable generation can be predicted day-ahead. For the feed-in renewable, we will discuss it with feed-in tariff in Section \ref{sec:model}-B.

\subsubsection{Energy Storage}
We assume that prosumer $i$ has an energy storage unit with a capacity of $E_i^{\text{B}}$. 
We let $e_i^{\text{B}}[t]$, $p_i^{\text{CH}}[t]$, and $p_i^{\text{DIS}}[t]$ denote the energy storage level, the amount of charging power and discharging power, respectively. The energy storage operation should satisfy the following constraints:
    \begin{align}
            e_i^{\text{B}}[t] {=} e_i^{\text{B}}[t{-}1] {+} \eta_i^{\text{CH}} p_i^{\text{CH}}[t] {-} \frac{1}{\eta_i^{\text{DIS}}} p_i^{\text{DIS}}[t], \forall i {\in} \mathcal{N}, t {\in} \mathcal{T}, \label{constraint-load6} \\
            \underline{\alpha}_i^{\text{B}} E_i^{\text{B}} \leq e_i^{\text{B}}[t] \leq \overline{\alpha}_i^{\text{B}} E_i^{\text{B}}, ~ \forall i \in \mathcal{N}, t \in \mathcal{T}, \label{constraint-load7} \\
            0 \leq  p_i^{\text{CH}}[t] \leq P_i^{\text{CH}}, ~ \forall i \in \mathcal{N}, t \in \mathcal{T}, \label{constraint-load8} \\
            0 \leq  p_i^{\text{DIS}}[t] \leq P_i^{\text{DIS}}, ~ \forall i \in \mathcal{N}, t \in \mathcal{T}, \label{constraint-load9}
    \end{align}
in which $\eta_i^{\text{CH}} \in [0,1]$ and $\eta_i^{\text{DIS}} \in (0,1]$ are the efficiencies of charging and discharging. Constraints \eqref{constraint-load7}-\eqref{constraint-load9} specify the box constraints for the energy storage level, the amount of charging power and discharging power.
For example, $\underline{\alpha}_i^{\text{B}}$ and $\overline{\alpha}_i^{\text{B}}$ denote the minimum and maximum fraction of the energy storage capacity for the feasible range of the energy storage level. Similarly, $P_i^{\text{CH}}$ and $P_i^{\text{DIS}}$ denote the maximum amount of charging power and discharging power, respectively. 

The energy storage operation incurs degradation, and we model its cost as
    \begin{align}
            \mathcal{C}_i^{\text{B}} = \beta_i^{\text{B}} \sum\nolimits_{t\in \mathcal{T}} \left( p_i^{\text{CH}}[t] + p_i^{\text{DIS}}[t] \right), \nonumber
    \end{align}
which is a function of charging and discharging power. Note that $\beta_i^{\text{B}}$ is the cost coefficient.

\subsection{Interaction of Distributed Energy Resources}
Prosumers can interact with each other to trade energy using their renewable and energy storage. Also, prosumers can feed renewable back to the grid and perform demand response using flexible loads. 

We first model the energy trading among prosumers. For a pair of prosumers $i$ and $j$, where $j \in \mathcal{N} \backslash i$, they can trade energy denoted as $p_{i,j}^{\text{ET}}[t]$. Note that $p_{i,j}^{\text{ET}}[t] >0$ indicates that prosumer $i$ sells energy to $j$; otherwise, $p_{i,j}^{\text{ET}}[t] <0$ indicates that prosumer $i$ purchases energy from $j$. We assume that the power loss during the energy trade is negligible. Thus, the energy trading for any pair $i$ and $j$ satisfies the following clearing constraints:
\begin{align}
    p_{i,j}^{\text{ET}}[t] + p_{i,j}^{\text{ET}}[t] = 0,&~\forall t \in \mathcal{T},~\forall i \in \mathcal{N},~\forall j \in \mathcal{N} \backslash i. \label{constraint-load10}
\end{align}

A feed-in tariff (FIT) provides rewards to FIT-eligible renewable generators to feed electricity back to the grid and thus serves as a policy instrument to promote the adoption of renewable energy. We denote the FIT rate as $\pi^{\text{FIT}}[t]$ for all the prosumers. Each prosumer $i$ decides how much renewable to sell to the grid, which is denoted as $p_i^{\text{FIT}}[t]$. Therefore, we have the following constraints for feed-in energy:
\begin{align}
    0 \leq p_i^{\text{FIT}}[t] \leq P_i^{\text{RE}}[t] - p_i^{\text{RE}}[t], &~ \forall i \in \mathcal{N}, \forall t \in \mathcal{T}, \label{constraint-load11} 
\end{align}
in which the feed-in renewable energy $p_i^{\text{FIT}}[t]$ is non-negative and no greater than the available renewable energy excluding the self-consumed renewable, i.e., $P_i^{\text{RE}}[t] - p_i^{\text{RE}}[t]$.

For the feed-in renewable energy, prosumer $i$ can earn a FIT revenue $\mathcal{R}_i^{\text{FIT}}$ as
\begin{align}
   \mathcal{R}_i^{\text{FIT}} = \sum\nolimits_{t\in \mathcal{T}} \pi^{\text{FIT}}[t] p_i^{\text{FIT}}[t]. \nonumber
\end{align}

The demand response program is another service that prosumers can provide and earn extra revenue. The grid operator usually sends DR requests during peak hours to ask prosumers to reduce their load. 
Prosumers can decide whether or not and how much load they can reduce, denoted as $p_i^{\text{DR}}[t]$. Once prosumers respond to the DR requests and perform load reduction, they can receive rewards at a unit rate of $\pi^{\text{DR}}[t]$ from the grid. The revenue from DR $\mathcal{R}_i^{\text{DR}}$ is written as
\begin{align}
   \mathcal{R}_i^{\text{DR}} = \sum_{t \in \mathcal{T}} \pi^{\text{DR}}[t] p_i^{\text{DR}}[t]. \nonumber
\end{align}

The prosumers' load reduction only counts their grid purchase and thus is bounded by its scheduled grid power purchase $p_i^{\text{G}}[t]$. Thus, we have the following constraints for the load reduction $p_i^{\text{DR}}[t]$:
\begin{align}
    0 \leq p_i^{\text{DR}}[t] \leq p_i^{\text{G}}[t], &~ \forall i \in \mathcal{N}, t \in \mathcal{T}. \label{constraint-load13} 
\end{align}

\subsection{Problem Formulation}
All the prosumers need to balance their supply and demand and aim to minimize the total costs. Specifically, prosumer $i$ needs to satisfy the 
the following balance constraints:
    \begin{equation}
        \begin{aligned}
            & p_i^{\text{AC}}[t] + p_i^{\text{S}}[t] + p_i^{\text{CH}}[t] + \sum\nolimits_{j \in \mathcal{N} \backslash i} p_{i,j}^{\text{ET}}[t] \\
            & =p_i^{\text{RE}}[t] + p_i^{\text{G}}[t] - p_i^{\text{DR}}[t] + p_i^{\text{DIS}}[t] , ~ \forall t \in \mathcal{T}, \label{constraint-load14}
        \end{aligned}
    \end{equation}
in which the left-hand side is the total demand of prosumer $i$, and the right-hand side is the total realized supply of $i$, all in time slot $t$.

For the simplicity of the notations, we define $\bm{p}_i^{\text{G}} = \{ p_i^{\text{G}}[t], \forall t\}$, $\bm{p}_i^{\text{AC}} = \{ p_i^{\text{AC}}[t], \forall t\}$, $\bm{\tau}_i^{\text{IN}} = \{ \tau_i^{\text{IN}}[t], \forall t\}$, $\bm{p}_i^{\text{S}} = \{ p_i^{\text{S}}[t], \forall t\}$, $\bm{p}_i^{\text{CH}} = \{ p_i^{\text{CH}}[t], \forall t\}$, $\bm{p}_i^{\text{DIS}} = \{ p_i^{\text{DIS}}[t], \forall t\}$, $\bm{p}_i^{\text{ET}} = \{ p_{i,j}^{\text{ET}}[t], \forall t,\forall j\}$, $\bm{p}_i^{\text{FIT}} = \{ p_i^{\text{FIT}}[t], \forall t\}$, and $\bm{p}_i^{\text{DR}} = \{ p_i^{\text{DR}}[t], \forall t\}$. We consider optimally coordinating DERs and aim to minimize the total cost of DERs. Therefore, we formulate the following DERs Cost Minimization problem.

\noindent\textbf{DCM}: DERs Cost Minimization
    \begin{equation*}
        \begin{aligned}
            &\text{minimize} {}&&{} \sum_{i\in\mathcal{N}} C_i^{\text{H}}(\bm{p}_i^{\text{G}},\bm{p}_i^{\text{AC}},\bm{\tau}_i^{\text{IN}},\bm{p}_i^{\text{S}},\bm{p}_i^{\text{CH}},\bm{p}_i^{\text{DIS}}) \\
            &&& -  \sum_{i\in\mathcal{N}} \left( R_i^{\text{ET}}(\bm{p}_i^{\text{ET}}) 
            + R_i^{\text{FIT}}(\bm{p}_i^{\text{FIT}})
            + R_i^{\text{DR}}(\bm{p}_i^{\text{DR}})
            \right) \\
            &\text{subject to} {}&&{} 
            \eqref{constraint-load1}-\eqref{constraint-load14}\\
            &\text{variables:} {}&&{}
            \{ \bm{p}_i^{\text{G}},\bm{p}_i^{\text{AC}},\bm{\tau}_i^{\text{IN}},\bm{p}_i^{\text{S}},\bm{p}_i^{\text{CH}},\bm{p}_i^{\text{DIS}},
            \bm{p}_i^{\text{ET}}, \bm{p}_i^{\text{FIT}},\bm{p}_i^{\text{DR}}\}.
        \end{aligned} 
    \end{equation*}

Note that the centralized method is not available for solving Problem \textbf{DCM}, as the centralized solution requires prosumers to reveal private information and thus causes serious privacy concerns. Therefore, we will develop a privacy-persevering distributed optimization algorithm to solve Problem \textbf{DCM} and discuss how to implement the algorithm as a smart contract working on a blockchain system in Section~\ref{sec:algorithm}.

\section{Distributed Algorithm on Blockchain}\label{sec:algorithm}
This section will present the design of a distributed algorithm to coordinate DERs, which can be implemented as a smart contract working on the blockchain.  

\subsection{Distributed Algorithm}\label{alg}
To preserve the prosumers' privacy and solve the optimal solution to Problem \textbf{DCM}, we design a distributed optimization algorithm that can be implemented as a smart contract on a blockchain system. 

The ADMM method \cite{boyd2011distributed} provides a distributed solution method, which has been used in energy trading \cite{wang2016incentivizing} for its good convergence and scalability. We first introduce $\hat{\bm{p}}_i^{\text{ET}} {=} \{\hat{p}_{i,j}^{\text{ET}}[t], \forall t,\forall j\}$ as the auxiliary variables for the energy trading decisions $\bm{p}_i^{\text{ET}}$. To replace \eqref{constraint-load10}, we introduce equivalent constraints as
    \begin{align}
        \hat{p}_{i,j}^{\text{ET}}[t] &= p_{i,j}^{\text{ET}}[t], \forall j {\in} \mathcal{N} \backslash i,\forall i {\in} \mathcal{N},\forall t {\in} \mathcal{T}, \label{constraint-auxiliary1}\\
        \hat{p}_{i,j}^{\text{ET}}[t] {+} \hat{p}_{j,i}^{\text{ET}}[t] &= 0, \forall j {\in} \mathcal{N} \backslash i,\forall i {\in} \mathcal{N},\forall t \in \mathcal{T}. \label{constraint-auxiliary2}
    \end{align}
    
We define the dual variables $\bm{\lambda}_{i} {=} \{ \lambda_{i,j}[t],\forall j {\in} \mathcal{N} \backslash i,t {\in} \mathcal{T}\}$ for constraints \eqref{constraint-auxiliary1} and decompose Problem \textbf{DCM} into the following two tasks executed by prosumers and a smart contract on the blokchain. \\

\noindent\textbf{PLT$_i$}: Prosumer $i$'s Local Task
    \begin{equation*}
        \begin{aligned}
            &\text{minimize} {}&&{} \sum_{i\in\mathcal{N}} C_i^{\text{H}}(\bm{p}_i^{\text{G}},\bm{p}_i^{\text{AC}},\bm{\tau}_i^{\text{IN}},\bm{p}_i^{\text{S}},\bm{p}_i^{\text{CH}},\bm{p}_i^{\text{DIS}}) \\
            &&& -  \sum_{i\in\mathcal{N}} \left( R_i^{\text{ET}}(\bm{p}_i^{\text{ET}}) 
            + R_i^{\text{FIT}}(\bm{p}_i^{\text{FIT}})
            + R_i^{\text{DR}}(\bm{p}_i^{\text{DR}})
            \right) \\
          &&& {+}  \sum_{t\in\mathcal{T}} \sum_{j \in \mathcal{N} \backslash i}
        \left[ \frac{\rho}{2} \left( \hat{p}_{i,j}^{\text{ET}}[t] {-} p_{i,j}^{\text{ET}}[t] \right)^{2} 
        {-} \lambda_{i,j}[t]  p_{i,j}^{\text{ET}}[t] \right] \\
            &\text{subject to} &&
            \eqref{constraint-load1}-\eqref{constraint-load9}, \eqref{constraint-load11}-\eqref{constraint-load14},\eqref{constraint-auxiliary2}\\
            &\text{variables:} &&
            \bm{p}_i^{\text{G}},\bm{p}_i^{\text{AC}},\bm{\tau}_i^{\text{IN}},\bm{p}_i^{\text{S}},\bm{p}_i^{\text{CH}},\bm{p}_i^{\text{DIS}},\bm{p}_i^{\text{ET}}, \bm{p}_i^{\text{FIT}},\bm{p}_i^{\text{DR}}.
        \end{aligned} 
    \end{equation*}
    
Prosumer $i$ determines the optimal energy schedule and trading for DERs in Task \textbf{PLT$_i$}. Then the prosumers call the smart contract in Task \textbf{SCT} to get the updated dual variables $\bm{\lambda}_{i}$ and auxiliary variables $\hat{\bm{p}}_i^{\text{ET}}$ to solve Problem \textbf{DCM} iteratively.

After receiving prosumers' energy trading decisions $\bm{p}_i^{\text{ET}}$, the smart contract solves the following optimization task.\\

\noindent\textbf{SCT}: Smart Contract Task
    \begin{equation*}
        \begin{aligned}
        &\text{minimize} && \sum_{t\in\mathcal{T}} \sum_{i\in\mathcal{N}} \sum_{j \in \mathcal{N} \backslash i}  
        \Big\{ \frac{\rho}{2} \left( \hat{p}_{i,j}^{\text{ET}}[t] - p_{i,j}^{\text{ET}}[t] \right)^{2} \\
        &&&  \quad\quad\quad\quad + \lambda_{i,j}[t] \hat{p}_{i,j}^{\text{ET}}[t] \Big\} \\
        &\text{subject to} &&
        \text{\eqref{constraint-auxiliary2}} \\
        &\text{variables:} &&
        \{ \hat{\bm{p}}_i^{\text{ET}},~ \forall i \in \mathcal{N} \},
        \end{aligned} 
    \end{equation*}
in which, the smart contract updates the dual variables $\boldsymbol{\lambda}_i$ and auxiliary variables $\boldsymbol{\hat{p}}_i^{\text{ET}}$ and broadcasts to prosumers. Specifically, Task \textbf{SCT} solves the optimal auxiliary variables as
        \begin{align}
            \begin{split}
                \hat{p}_{i,j}^{\text{ET}}[t]
                {=} \frac{\rho \left( p_{i,j}^{\text{ET}}[t] - p_{j,i}^{\text{ET}}[t] \right) - \left( \lambda_{i,j}[t] - \lambda_{j,i}[t] \right) }{2 \rho}, \label{updateenergy}
            \end{split}
        \end{align}
and updates the dual variables as
    \begin{align}
        \lambda_{i,j}[t] \leftarrow \lambda_{i,j}[t] + \rho \left( \hat{p}_{i,j}^{\text{ET}}[t] - p_{i,j}^{\text{ET}}[t] \right). \label{updatelambda}
    \end{align}

\subsection{Blockchain for the Coordination of DERs}

In our work, we employ blockchain technology to facilitate the coordination of DERs for three purposes. First, we adopt the blockchain to implement an open and verifiable DER energy coordination platform. Unlike the conventional centralized energy coordination, the blockchain provides a trustable computing machine that can execute the DER coordination algorithm written in a smart contract. Therefore, the use of the blockchain removes the need for a centralized coordinator and guarantees the correctness of the coordination algorithm. Second, the blockchain network can be used as a secure and robust data communication network to exchange information among prosumers. Third, the blockchain also provides an efficient payment tool for energy trading and service rewards.

\begin{figure}[!t]
    \centering
    \includegraphics[width=8.5cm]{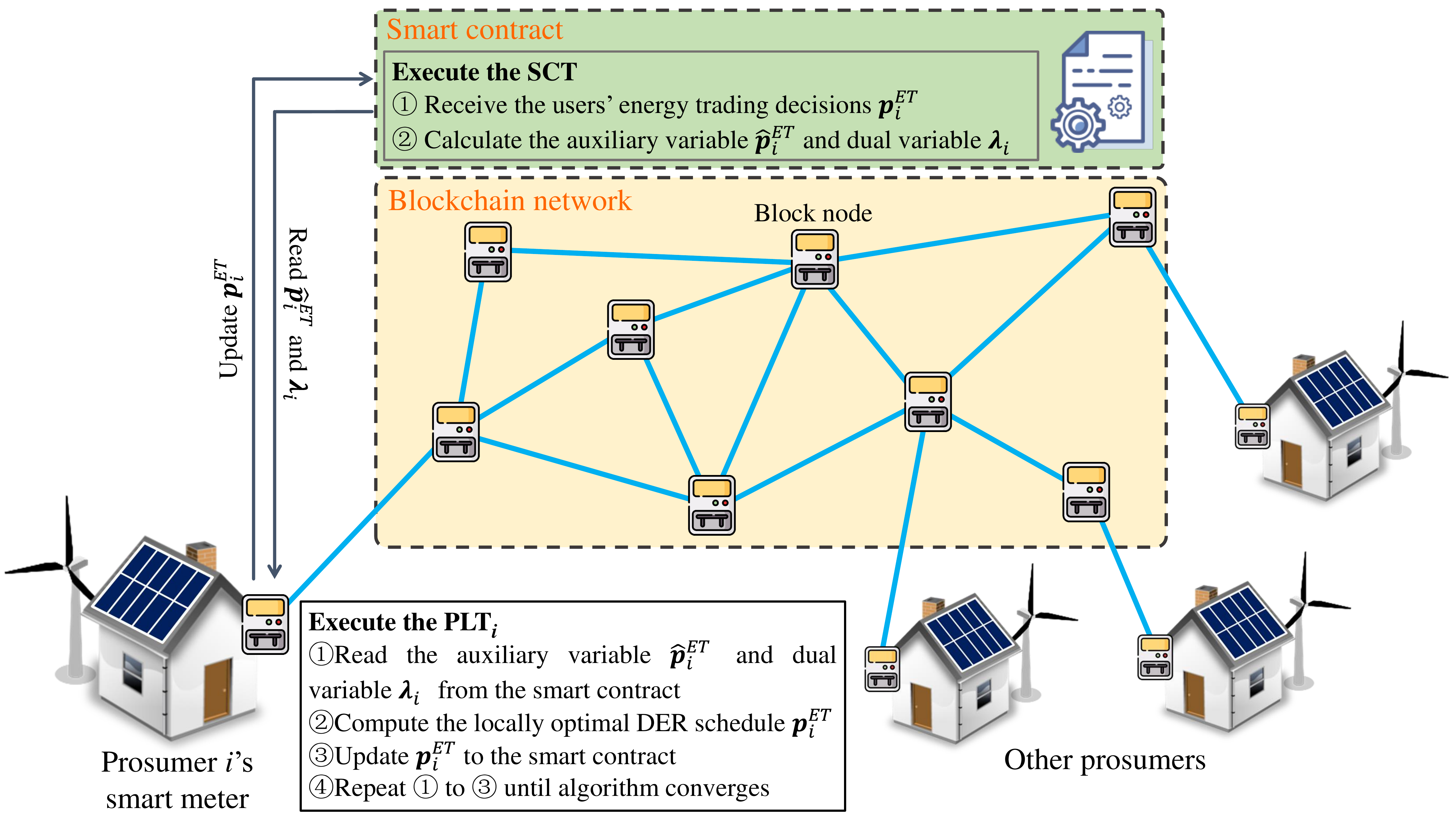}
    \caption{The process of the distributed DER coordination algorithm on blockchain.}
    \label{f:process}
\end{figure}

For the underlying blockchain system, we let the users' smart meters be the blockchain nodes to form the blockchain network, as shown in Fig.~\ref{f:process}. Modern smart meters are equipped with embedded systems that support embedded operating systems (e.g., Ubuntu Core) and application software, including the blockchain software. The prosumers' smart meters can connect to the blockchain network via the wireless communication link such as LoRa and 5G Narrowband IoT. The connected smart meters form a peer-to-peer network that supports data communication and the operation of the blockchain.

We choose the Ethereum blockchain with proof-of-authority (PoA) consensus protocol (Clique) based on the following considerations in this work. First, Ethereum is an open and mature blockchain project supported by most of the operating systems and hardware. Second, the consensus protocol, which is used to synchronized the states of all the blockchain nodes, has a critical impact on the performance of the blockchain. We select the PoA consensus rather than proof-of-work (PoW) because the computational complexity of PoW is prohibitively high for smart meters' hardware resources. Third, Ethereum supports the smart contract for the users to implement their function in the Solidity programming language.

The blockchain nodes in Fig.~\ref{f:process} can be divided into two categories: \emph{PoA nodes} and \emph{normal nodes}. The normal nodes are the normal prosumers that can send transactions and interact with the smart contract. The normal nodes can also access the data on the blockchain and verify the execution of the transactions. Among the normal nodes, some are selected as the PoA nodes that participate in the consensus process of the blockchain. Specifically, a group of PoA nodes forms a committee to receive the transactions, execute smart contracts, and package them to generate a new block in a round-robin manner. The PoA nodes can also vote to add a node into or remove a node out of the committee.  

As discussed in Section~\ref{alg}, we implement the \textbf{SCT} part of the distributed algorithm in the smart contract on the blockchain. Since the execution of the smart contract is transparent and verifiable, the result of the \textbf{SCT} is guaranteed to be correct and trustable. Specifically, the smart contract in Fig.~\ref{f:process} implements three functions. The first function is to let the prosumers update their energy trading decisions $\bm{p}_{i}^{\text{ET}}$. The second function is to solve the problem of \textbf{SCT} as in Eq.~\eqref{updateenergy} and Eq.~\eqref{updatelambda}. The third function is to let the prosumers read the latest value of dual variables $\boldsymbol{\lambda}_i$ and auxiliary variables $\boldsymbol{\hat{p}}_i^{\text{ET}}$. In each iteration of the distributed algorithm, the prosumer $i$ first read the value of $\boldsymbol{\lambda}_i$ and $\boldsymbol{\hat{p}}_i^{\text{ET}}$ from the smart contract, then solve the problem of \textbf{PLT$_i$} to compute $\bm{p}_{i}^{\text{ET}}$, and finally update this value to the smart contract. The smart contract automatically executes the second function when all the trading decisions are updated. This algorithm iterates until the results converge to the optimal coordination schedule.

\section{Numerical Results}\label{sec:results}
To evaluate the performance of our design, we simulate the blockchain-based DER coordination system with a distributed algorithm using real-world data \cite{wang2015joint, pecan}. Due to the page limit, we do not include the implementation detail.

\begin{figure}[!b]
    \centering
    \includegraphics[width=8.5cm]{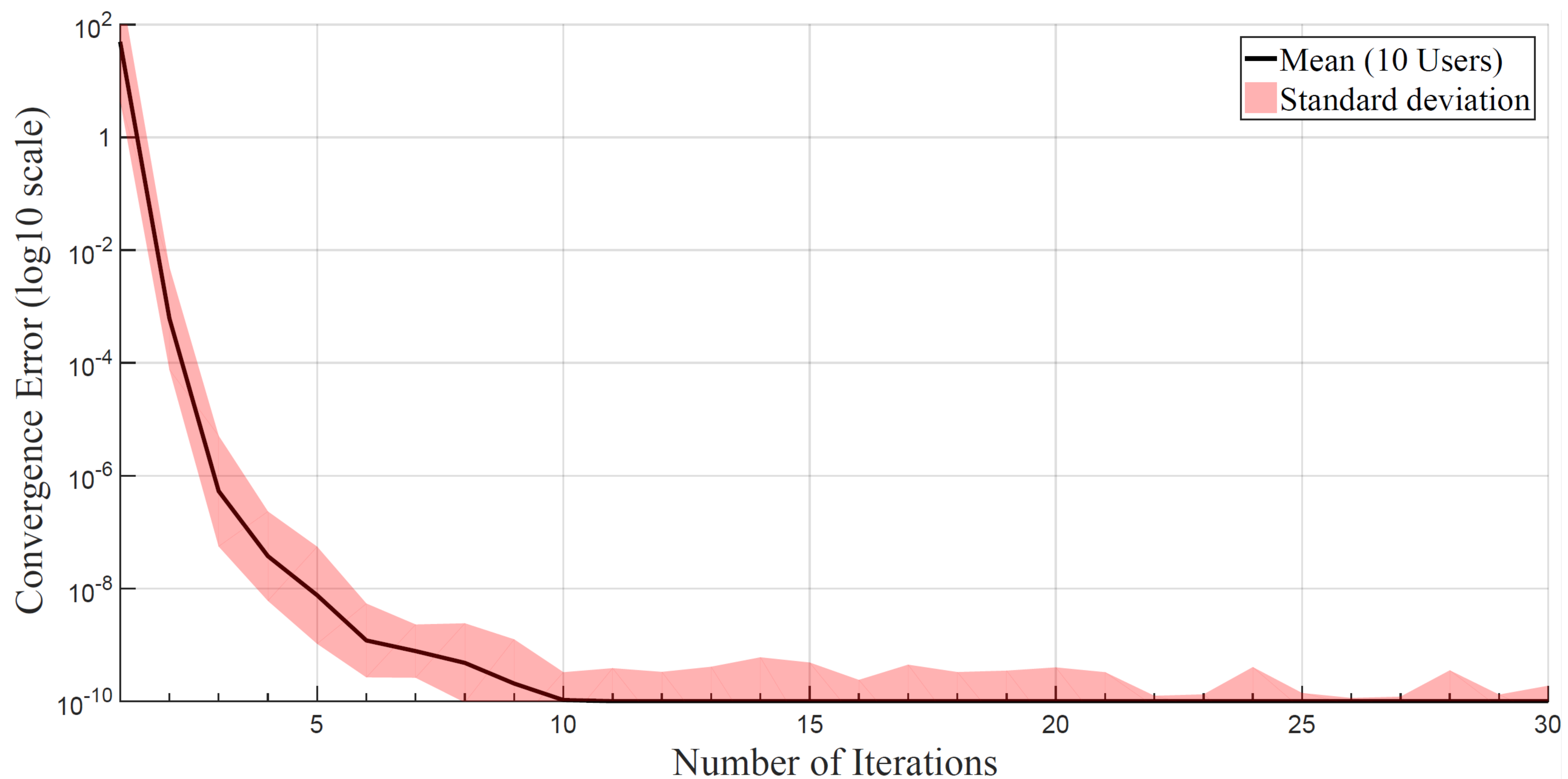}
    \caption{The convergence of the proposed distributed optimization algorithm.}
    \label{f:converge}
\end{figure}

\begin{figure}[!b]
    \centering
    \includegraphics[width=8.5cm]{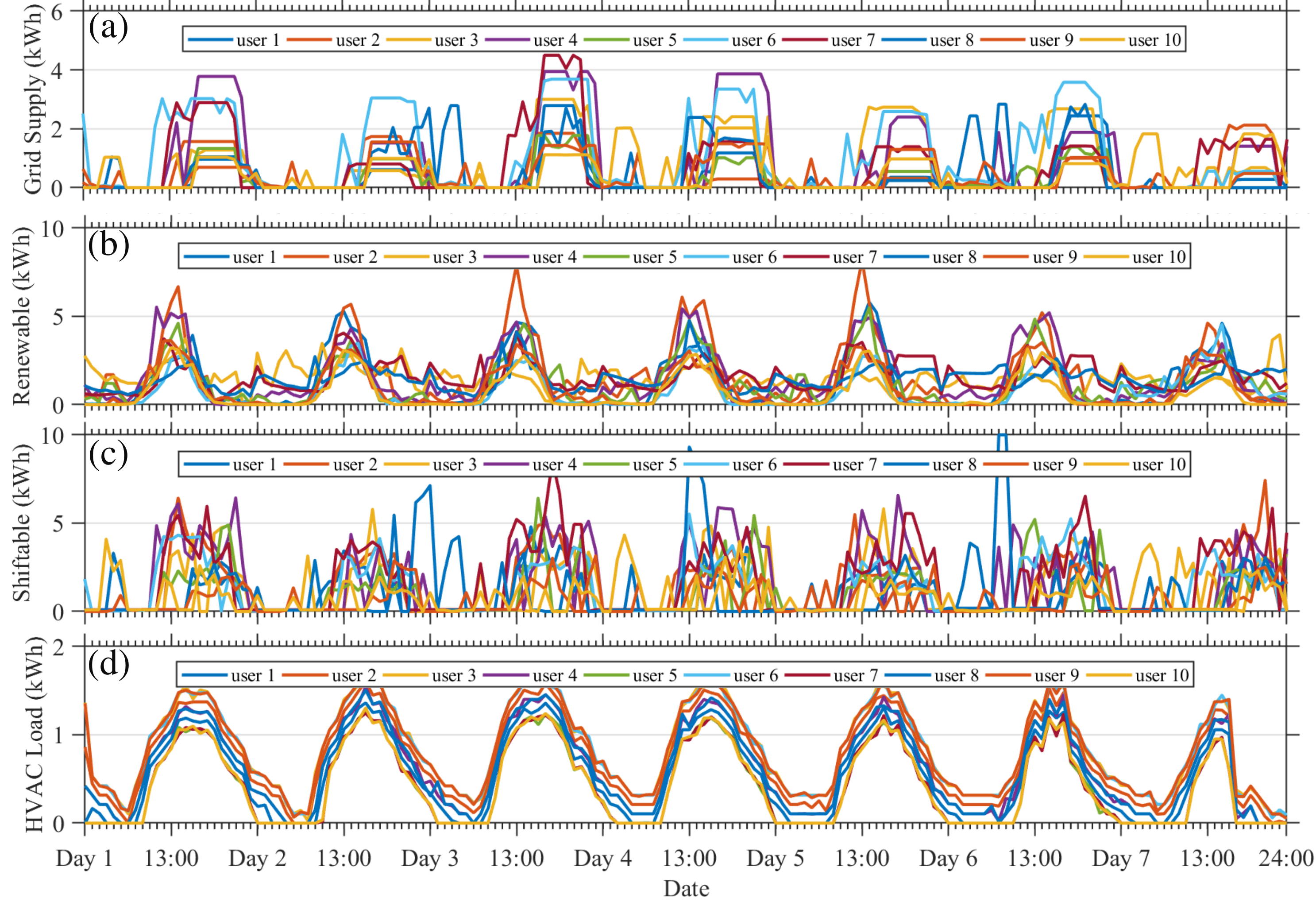}
    \caption{The optimal energy schedule of DERs including (a) grid supply, (b) renewable energy supply, (c) shiftable load, and (d) HVAC load.}
    \label{f:schedule}
\end{figure}

First, we evaluate the convergence of the distributed optimization algorithm in Section~\ref{sec:algorithm} with $10$ prosumers. We see in Fig.~\ref{f:converge} that the algorithm converges within $10$ iterations to keep the error below the preset error threshold. 

Second, we show the optimal energy schedule of $10$ prosumers over one week in Fig.~\ref{f:schedule}. Prosumers' grid supply and renewable profiles in \ref{f:schedule}(a)-(b) are complementary to each other, as prosumers first use their local renewables and then use grid power if there is a deficit in renewable generation. The optimal scheduling of shiftable and HVAC loads in \ref{f:schedule}(c)-(d) exhibits a similar trend because they have similar shiftable appliances and face the same outdoor temperature. But the shiftable and HVAC loads still show some diversity as prosumers' operational preferences of their DERs are different. 

\begin{figure}[!t]
    \centering
    \includegraphics[width=8.5cm]{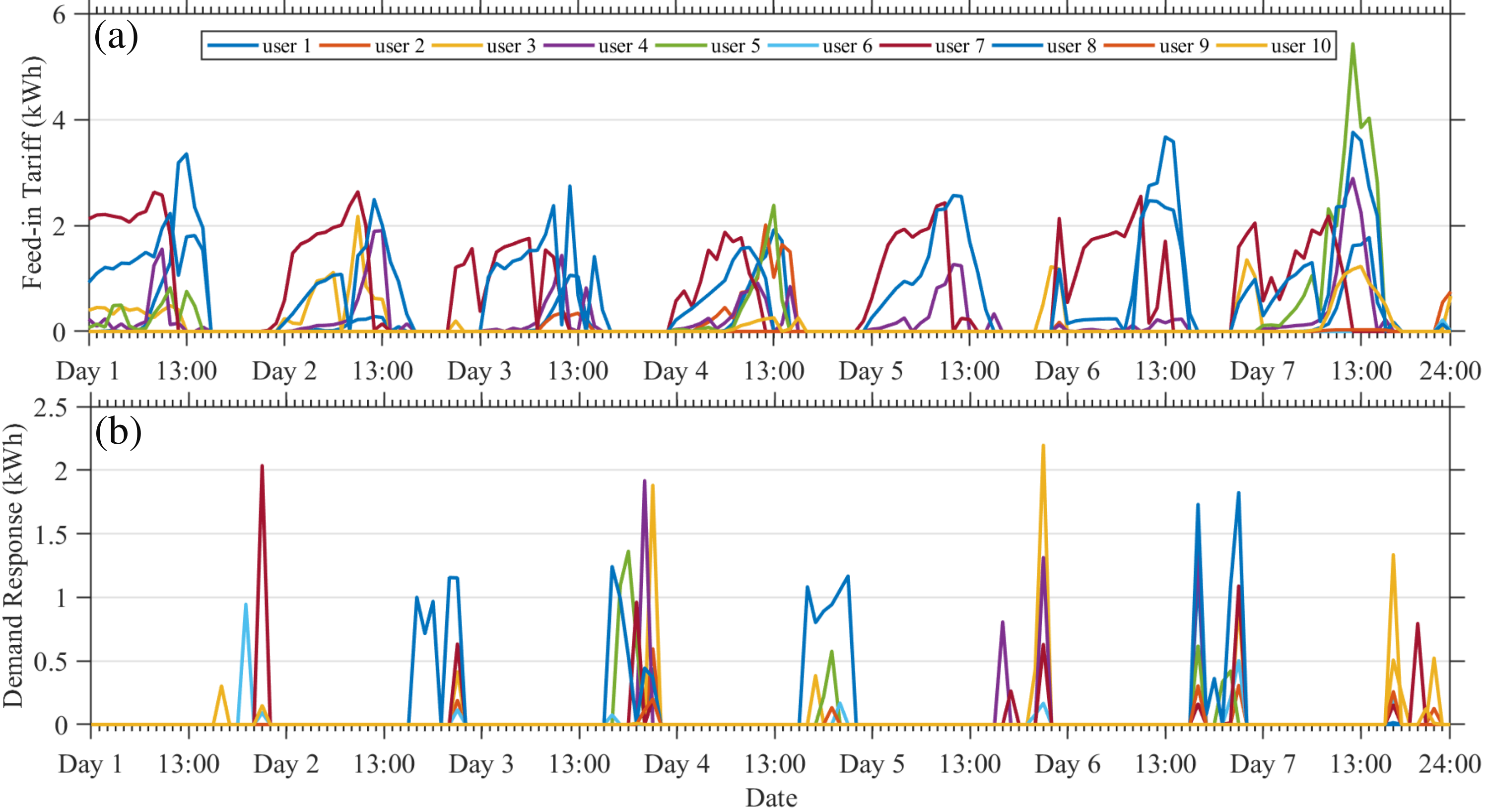}
    \caption{The service schedule of (a) the feed-in energy and (b) the demand response.}
    \label{f:vpp}
\end{figure}

Third, Fig.~\ref{f:vpp} illustrates the optimized feed-in energy and demand response. We see from Fig.~\ref{f:vpp}(a) that prosumers sell extra renewable generations to the grid when the system has a low demand. In Fig.~\ref{f:vpp}(b), prosumers perform differently in demand response as they have different local generation and load profiles, and some of them do not have the capacity to respond to demand response signals in the peak hours. 

\begin{figure}[!t]
    \centering
    \includegraphics[width=8.5cm]{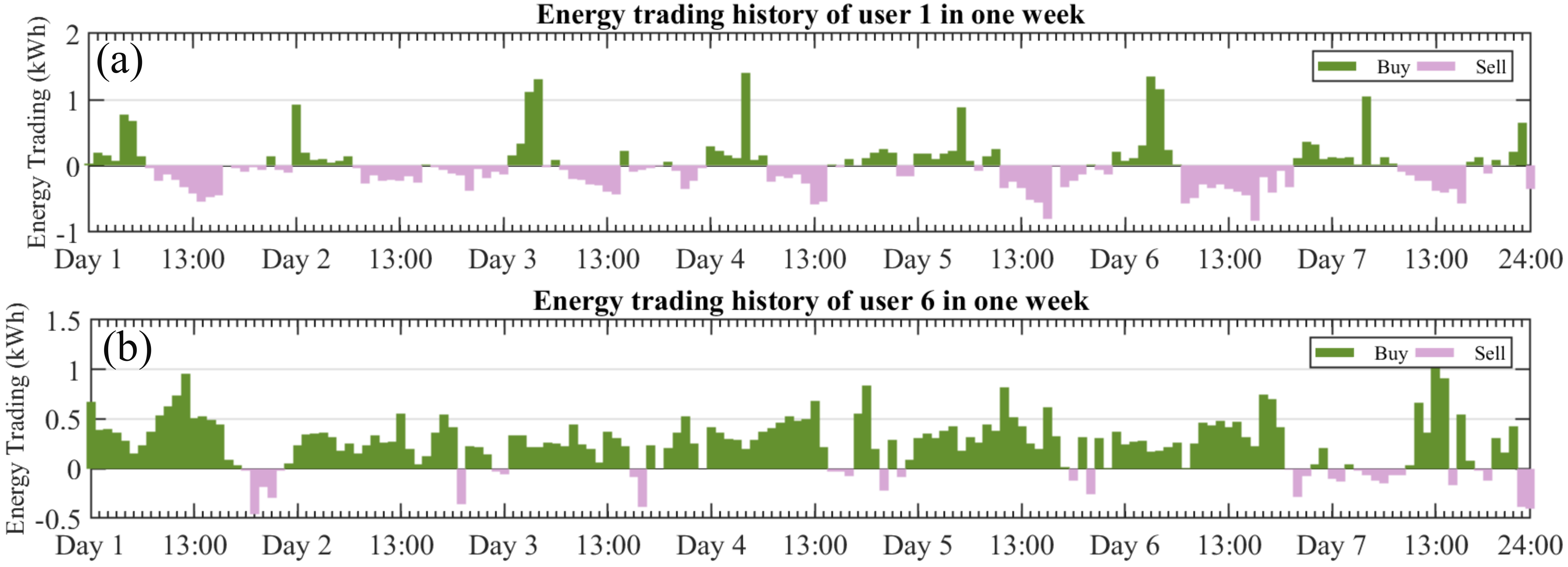}
    \caption{The energy trading records of two typical users: (a) user 1 and (b) user 6.}
    \label{f:et}
\end{figure}

Fourth, Fig.~\ref{f:et} shows the optimal energy trading of two typical prosumers. We see that prosumers 1 and 6 are two different types, as prosumer 1 has higher renewable generations and sells a lot to others, while prosumer 6 lacks local renewables and needs to buy energy from other prosumers most of the time. The blockchain-based DER coordination system helps the prosumers to trade energy with each other and interact with the grid to sell energy or perform demand response, which significantly improves the efficiency of the system. 

\section{Conclusion}\label{sec:conclusion}
This paper developed a blockchain system for the coordination of distributed energy resources. We modeled energy trading, feed-in energy, and demand response for various DERs. Given prosumers' independence, we designed a distributed optimization algorithm to coordinate DERs to minimize their costs. More importantly, we developed a blockchain system with a smart contract to execute the distributed optimization algorithm for DERs to facilitate the trustable and efficient integration of DERs. We validated our design by numerical simulations using real-world data.

For our future work, we will 1) model more services that DERs can provide, 2) consider network constraints for DERs, and 3) implement the blockchain system on the Internet of things devices. 

\bibliographystyle{IEEEtran}
\bibliography{ref}

\end{document}